\documentclass[11pt,a4paper,oneside,parskip=half,abstract=true]{scrartcl}

\usepackage[a4paper,margin=1in]{geometry}
\usepackage[T1]{fontenc}
\usepackage[utf8]{inputenc}
\usepackage{lmodern}
\usepackage{microtype}
\usepackage{xurl}
\usepackage[htt]{hyphenat}

\usepackage{amsmath}
\usepackage{amssymb}
\usepackage{graphicx}
\graphicspath{{./}{figures/}{images/}}
\usepackage{algorithm}
\usepackage{algorithmicx}
\usepackage{algpseudocode}
\usepackage{booktabs}
\usepackage{multirow}
\usepackage{array}
\usepackage{siunitx}
\usepackage{float}
\usepackage[section]{placeins}

\usepackage{pgfplots}
\usepgfplotslibrary{groupplots}
\pgfplotsset{compat=1.18}
\usepackage[table,dvipsnames]{xcolor}

\usepackage[square,numbers]{natbib}
\usepackage[breaklinks=true,colorlinks=true,linkcolor=black,citecolor=black,urlcolor=blue]{hyperref}
\hypersetup{
  pdftitle={GLM-5 Serving Parameter Tuning for OpenClaw: Single-Deployment MaaS Inference Optimization for Long-Context Agent Workloads},
  pdfauthor={Minjie Hua, Ning Wang, Peijun Yang, Kai Wang, and Shiguo Lian},
  pdfsubject={Technical Report},
  pdfkeywords={GLM-5, OpenClaw, SGLang, MaaS, long-context inference, tensor parallelism, pipeline parallelism, chunked prefill}
}

\usepackage{scrlayer-scrpage}
\clearpairofpagestyles
\setkomafont{title}{\sffamily\bfseries\LARGE}
\setkomafont{subtitle}{\sffamily\large}
\setkomafont{section}{\sffamily\bfseries\Large}
\setkomafont{subsection}{\sffamily\bfseries\large}
\setkomafont{caption}{\small}
\setkomafont{captionlabel}{\bfseries}
\RedeclareSectionCommand[beforeskip=1.2\baselineskip,afterskip=.55\baselineskip]{section}
\RedeclareSectionCommand[beforeskip=.9\baselineskip,afterskip=.4\baselineskip]{subsection}

\renewenvironment{abstract}{%
  \par\bigskip\begingroup\small
  \begin{center}\sffamily\bfseries\normalsize\abstractname\end{center}
  \noindent\ignorespaces
}{%
  \par\endgroup\bigskip
}

\subject{Technical Report}

\title{GLM-5 Serving Parameter Tuning for OpenClaw\\Single-Deployment MaaS Inference Optimization for Long-Context Agent Workloads}

\author{%
Minjie Hua, Ning Wang, Peijun Yang, Kai Wang, and Shiguo Lian\\
China Unicom
}
\date{\today}

\newcommand{\WorkloadAvgInputTokLow}{28000}
\newcommand{\WorkloadAvgInputTokHigh}{30000}
\newcommand{\WorkloadAvgOutputTok}{500}

\newcommand{\RecChunk}{3072}
\newcommand{\RecTP}{4}
\newcommand{\RecPP}{4}
\newcommand{\RecMaxRun}{24}

\newcommand{\LegacyChunk}{4096}
\newcommand{\LegacyTP}{4}
\newcommand{\LegacyPP}{4}
\newcommand{\LegacyMaxRun}{16}

\newcommand{\BaseChunk}{2048}
\newcommand{\BaseTP}{4}
\newcommand{\BasePP}{4}
\newcommand{\BaseMaxRun}{16}

\newcommand{\RecTotalTokTput}{9993.23}
\newcommand{\RecReqTput}{0.48}
\newcommand{\RecTTFTAvg}{6.69}

\newcommand{\RecLatPNineZero}{32.64}

\newcommand{\LegacyTotalTokTput}{7885.54}
\newcommand{\LegacyReqTput}{0.38}

\newcommand{\BaseTotalTokTput}{9029.64}
\newcommand{\BaseReqTput}{0.43}
\newcommand{\BaseTTFTAvg}{8.98}

\newcommand{\BaseLatPNineZero}{40.23}

\newcommand{\GainReqVsBase}{11.6\%}
\newcommand{\GainTotVsBase}{10.7\%}
\newcommand{\DropTTFTVsBase}{25.5\%}
\newcommand{\DropLatVsBase}{18.9\%}
\newcommand{\CostReqVsBase}{10.4\%}
\newcommand{\CostTokVsBase}{9.6\%}

\newcommand{\CostReqVsLegacy}{20.8\%}
\newcommand{\CostTokVsLegacy}{21.1\%}

\newcommand{\NormCostReqBase}{1.000}
\newcommand{\NormCostTokBase}{1.000}
\newcommand{\NormCostReqLegacy}{1.132}
\newcommand{\NormCostTokLegacy}{1.145}
\newcommand{\NormCostReqRec}{0.896}
\newcommand{\NormCostTokRec}{0.904}

\definecolor{ccblue}{RGB}{46,134,222}
\definecolor{ccgreen}{RGB}{39,174,96}
\definecolor{ccorange}{RGB}{243,156,18}
\definecolor{ccred}{RGB}{231,76,60}
\definecolor{ccpurple}{RGB}{142,68,173}
\definecolor{ccgray}{RGB}{90,98,104}

\newcolumntype{L}[1]{>{\raggedright\arraybackslash}p{#1}}

\begin{document}
\maketitle

\begin{center}
{\footnotesize huamj5@chinaunicom.cn, wangn85@chinaunicom.cn, yangpj16@chinaunicom.cn\\
wangk115@chinaunicom.cn, liansg@chinaunicom.cn}
\end{center}

\begin{abstract}
OpenClaw requests are dominated by long, tool-augmented prefixes: a sizable system prompt, accumulated conversation history, and tool outputs that are fed back into the context window \citep{openclaw_context,openclaw_system_prompt}. Under this workload shape---approximately \WorkloadAvgInputTokLow{} to \WorkloadAvgInputTokHigh{} input tokens per request and about \WorkloadAvgOutputTok{} output tokens---serving quality is governed jointly by throughput, time-to-first-token (TTFT), and tail latency rather than by raw short-prompt throughput alone \citep{sglang_bench_serving,vllm_anatomy_metrics}.

This report positions the tuning work inside a Model-as-a-Service (MaaS) multi-model inference optimization architecture. The broader architecture spans domain applications, MaaS platform performance optimization, multi-model orchestration, inference optimization, and compute-stack adaptation; the scope of this report is the \emph{Single-Node Optimization} block in the inference-optimization layer, where serving parameters such as chunked prefill, tensor parallelism (TP), pipeline parallelism (PP), and max-running-request concurrency are tuned for one GLM-5 serving deployment. In this report, \emph{Single-Node Optimization} is the architecture-block label, not a restriction to one physical server; the measured deployment uses the two-node, sixteen-GPU cluster described in Section~\ref{sec:experiments}.

Within the tested parameter space, the best measured configuration is \texttt{chunked-prefill-size=3072}, \texttt{tp=4}, \texttt{pp-size=4}, and \texttt{max-running-requests=24}. Relative to the conservative \texttt{2048/4/4/16} baseline, this setting improves request throughput from \BaseReqTput{} to \RecReqTput{} req/s and total token throughput from \BaseTotalTokTput{} to \RecTotalTokTput{} tok/s, while reducing TTFT average from \BaseTTFTAvg{} to \RecTTFTAvg{} seconds and latency P90 from \BaseLatPNineZero{} to \RecLatPNineZero{} seconds. Under a fixed hardware footprint, that corresponds to an estimated \CostReqVsBase{} lower serving cost per request and \CostTokVsBase{} lower serving cost per token. Relative to the \texttt{4096/4/4/16} comparison profile, normalized serving cost falls by about \CostReqVsLegacy{} per request and \CostTokVsLegacy{} per token.

Accordingly, we recommend \texttt{3072 / tp4 / pp4 / max24} as the default deployment profile for OpenClaw. The data also show that the optimum is not monotonic: larger chunk sizes (\texttt{4096}, \texttt{6144}) and deeper queueing (\texttt{32}, \texttt{48}) do not further improve this workload. The results indicate a workload-specific sweet spot rather than a simple ``larger is always better'' rule.
\end{abstract}

\section{Background and Problem Definition}
Optimizing serving parameters using ``raw model throughput'' (e.g., short prompts, single-turn, synthetic load) can be misleading for OpenClaw. OpenClaw constructs a per-run context that includes: (i) a system prompt describing tools/skills/runtime constraints, (ii) the full conversation history, and (iii) tool calls and tool outputs; all of these consume tokens inside the model context window \citep{openclaw_context,openclaw_system_prompt}. As a result, even when the \emph{user} prompt is brief, the effective model input can become very long, especially as multi-turn interactions accumulate and tool outputs (command output, file snippets, web/tool results) are reinjected \citep{openclaw_context}.

This yields three business-relevant characteristics for OpenClaw agent serving:

\textbf{Long system prompt and high context overhead.} Tool schemas and injected workspace content add non-trivial context cost each turn \citep{openclaw_system_prompt}. This increases the probability of ``long prefix'' requests even in routine usage.

\textbf{Multi-turn accumulation and tool-heavy traces.} Multi-round agent sessions are inherently prefix-growing: each turn appends assistant outputs and tool transcripts \citep{openclaw_context}. Long-prefix requests make prefill time and KV-cache pressure first-order performance factors \citep{glm5_arxiv}.

\textbf{Interactive stability over peak throughput.} In an agent product, user experience is driven by responsiveness (TTFT) and predictability (tail latency), not by maximizing requests-per-second at the cost of jitter and queueing \citep{vllm_anatomy_metrics,sglang_bench_serving}. Over-aggressive batching can turn the endpoint into a throughput-first queueing system, where some requests wait much longer before streaming begins.

Given this workload, three parameters become especially critical:

\textbf{Chunked prefill size.} SGLang exposes \texttt{--chunked-prefill-size} to control the maximum tokens per prefill chunk \citep{sglang_server_args}. This directly affects how long prefixes are partitioned, which in turn influences prefill efficiency, interference with decode, and observed TTFT under load.

\textbf{Tensor parallelism and pipeline parallelism.} \texttt{--tp-size} and \texttt{--pp-size} determine how the model computation is split across devices \citep{sglang_server_args}. TP trades compute locality for collective communication (as in Megatron-style tensor model parallelism), while PP introduces pipeline scheduling and potential ``bubble'' overhead that depends on stage balance and micro-batching \citep{megatron_lm,gpipe,vllm_parallelism_scaling}.

\textbf{Max running requests.} \texttt{--max-running-requests} caps the number of running requests and thereby constrains effective batch size and queue depth \citep{sglang_server_args}. It is a primary knob for throughput versus latency stability; SGLang guidance explicitly frames it as a control surface for memory safety and batching behavior \citep{sglang_hparam_tuning}.

\section{MaaS Multi-Model Inference Optimization Architecture}
\label{sec:maas_architecture}

\begin{figure}[!htbp]
\centering
\IfFileExists{architecture.png}{%
  \includegraphics[width=0.98\textwidth]{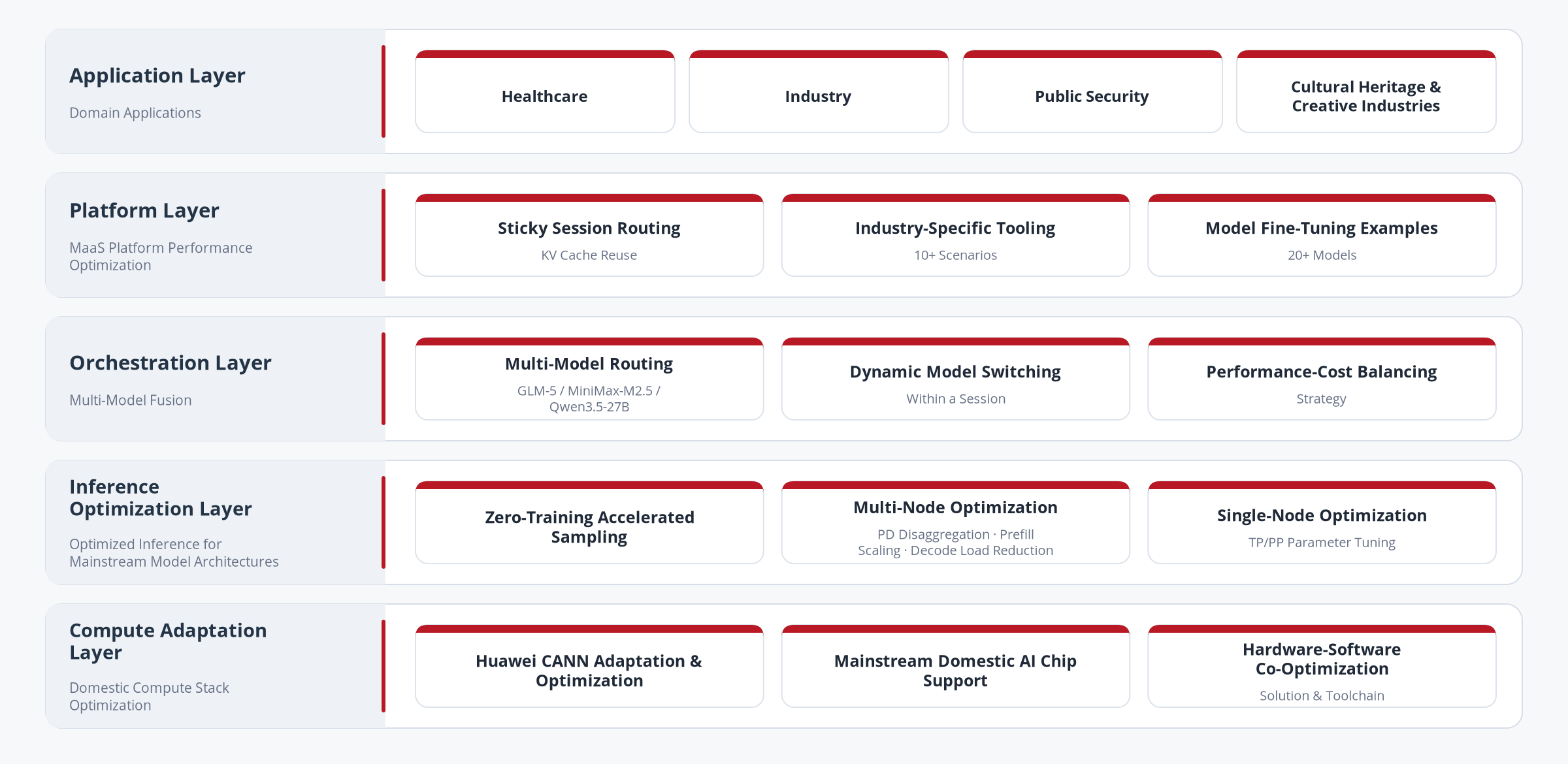}%
}{%
  \fbox{%
    \begin{minipage}{0.90\textwidth}
    \centering
    \vspace{1.6cm}
    \textbf{Architecture image placeholder}\\[0.5em]
    Expected file name: \texttt{architecture.png}\\
    Upload \texttt{architecture.png} to the same Overleaf directory as this \texttt{.tex} file.
    \vspace{1.6cm}
    \end{minipage}%
  }%
}
\caption{MaaS Multi-Model Inference Optimization Architecture. The architecture separates domain applications, MaaS platform performance optimization, multi-model orchestration, inference optimization, and compute adaptation. The experimental scope of this report corresponds to the Single-Node Optimization path, where TP/PP parameter tuning is evaluated for a GLM-5 serving deployment.}
\label{fig:maas_architecture}
\end{figure}

Figure~\ref{fig:maas_architecture} shows the MaaS Multi-Model Inference Optimization Architecture that frames this report. The architecture is organized as a five-layer stack: an application layer for domain-facing workloads, a platform layer for MaaS performance optimization, an orchestration layer for multi-model fusion, an inference optimization layer for request scheduling and model-serving efficiency, and a compute adaptation layer for accelerator- and toolchain-specific execution. This report focuses on the \textbf{Single-Node Optimization} block in the inference optimization layer, especially TP/PP parameter tuning and the adjacent serving controls that determine prefill behavior and request concurrency.

The \textbf{application layer} captures the domain workloads served by the MaaS platform. In the illustrated architecture, these workloads include healthcare, industry, public security, and cultural heritage/creative-industry applications. This layer is important because each domain can produce different prompt shapes, safety constraints, latency targets, and tool-use patterns. For OpenClaw, the dominant workload shape is a long-context agent session rather than a short single-turn prompt, so the serving profile must be optimized for large prefills and interactive streaming latency.

The \textbf{platform layer} represents MaaS platform performance optimization. Its mechanisms include sticky session routing for KV-cache reuse, industry-specific tooling, and model fine-tuning examples. Sticky session routing is especially relevant for long-context agents: when requests from the same session are routed consistently, repeated prefixes and accumulated session context can be reused more effectively, reducing avoidable recomputation and improving TTFT stability. The tooling and fine-tuning components provide the surrounding platform capabilities needed to specialize the MaaS service for multiple application domains.

The \textbf{orchestration layer} implements multi-model fusion. The architecture explicitly includes multi-model routing across models such as GLM-5, MiniMax-M2.5, and Qwen3.5-27B, dynamic model switching within a session, and performance--cost balancing. These functions decide which model should serve a request or a session segment. This report does not tune cross-model routing policies; instead, it holds model selection fixed to the GLM-5 serving path so that the impact of single-deployment serving parameters can be measured cleanly.

The \textbf{inference optimization layer} is the main technical layer for this study. It contains zero-training accelerated sampling, multi-node optimization, and single-node optimization. The multi-node branch covers mechanisms such as prefill/decode disaggregation, prefill scaling, and decode-load reduction. The single-node branch covers TP/PP parameter tuning, which is the direct focus of the present work. In addition to TP/PP, the experiments tune \texttt{chunked-prefill-size} and \texttt{max-running-requests}, because these parameters determine how long-prefix requests are partitioned, admitted, and batched before decoding begins.

The \textbf{compute adaptation layer} maps the optimized inference workload onto the available compute stack. The architecture includes Huawei CANN adaptation and optimization, mainstream domestic AI chip support, and hardware--software co-optimization through solution and toolchain work. Although the measurements in this report are collected on the fixed experimental hardware described later, the role of this layer is to ensure that the same optimization method can be adapted to different accelerator backends and deployment toolchains.

In this report, \emph{Single-Node Optimization} refers to optimizing one GLM-5 serving deployment path while holding the higher-level MaaS routing and multi-model orchestration policies constant. It should not be interpreted as a claim that the measured hardware footprint is limited to one physical server; rather, it identifies the architectural scope of the work: local serving-parameter tuning for a single model-serving path. The recommended \texttt{3072 / tp4 / pp4 / max24} profile should therefore be read as the best tested single-deployment configuration for the OpenClaw long-context workload under the measured environment.

\FloatBarrier

\section{Methodology}
This section defines the workload abstraction, evaluation metrics, and controlled-comparison protocol used to study serving-parameter sensitivity for OpenClaw.

\textbf{Benchmark workload and traffic abstraction.} The objective of the benchmark is not to maximize synthetic short-prompt throughput, but to approximate the request shape of \textbf{OpenClaw online traffic}, which is dominated by long system prompts, accumulated conversation history, and tool-returned context \citep{openclaw_context}. To reflect this property, the experiments use \textbf{SGLang \texttt{bench\_serving}} with a long-input synthetic profile rather than a short-context microbenchmark. A representative command uses \texttt{--dataset-name random}, \texttt{--random-input-len 30000}, \texttt{--random-output-len 500}, and \texttt{--random-range-ratio 0.2}, together with the same model and tokenizer path as the online deployment. Under this setup, the generated requests have an average input length of approximately \WorkloadAvgInputTokLow{}--\WorkloadAvgInputTokHigh{} tokens and an average output length of approximately \WorkloadAvgOutputTok{} tokens. This workload construction is a deliberate abstraction: it does not replay production traces token by token, but it preserves the dominant systems characteristic relevant to this study---\emph{very long prefills with moderate decode length}. That choice makes the benchmark suitable for answering the report's central question, namely which server parameters best balance throughput and latency for the OpenClaw long-context serving regime.

\textbf{Evaluation metrics and interpretation.} The current sweep records the following directly measured metrics from the benchmark output \citep{sglang_bench_serving}: (i) \emph{output token throughput} (tok/s), (ii) \emph{total token throughput} (tok/s), (iii) \emph{request throughput} (req/s), (iv) \emph{average TTFT}, (v) \emph{TTFT P90}, (vi) \emph{average TPOT}, and (vii) \emph{end-to-end latency P90}. Together, these metrics capture both system efficiency and user-perceived responsiveness. Throughput metrics quantify hardware utilization under sustained load, while TTFT and percentile latency characterize interactivity and stability under queueing pressure \citep{vllm_anatomy_metrics,anyscale_latency_metrics}. The analysis centers on fields directly reported by the benchmark output. Metrics such as average end-to-end latency, ITL, or P99 latency are not introduced unless they are explicitly available, which avoids overstating precision or fabricating unsupported comparisons.

\textbf{Methodological principle.} The study follows a controlled-comparison design: keep the deployment platform, model artifact, and software stack fixed; vary only the serving parameters of interest; and evaluate each candidate under a workload that reflects the long-context characteristics of OpenClaw. The study does not claim a universal optimum for GLM-5 serving. Rather, it identifies the most effective configuration \emph{within the tested parameter space} for a specific production-like workload running on a specific FP8 model artifact and a specific two-node H100 cluster.

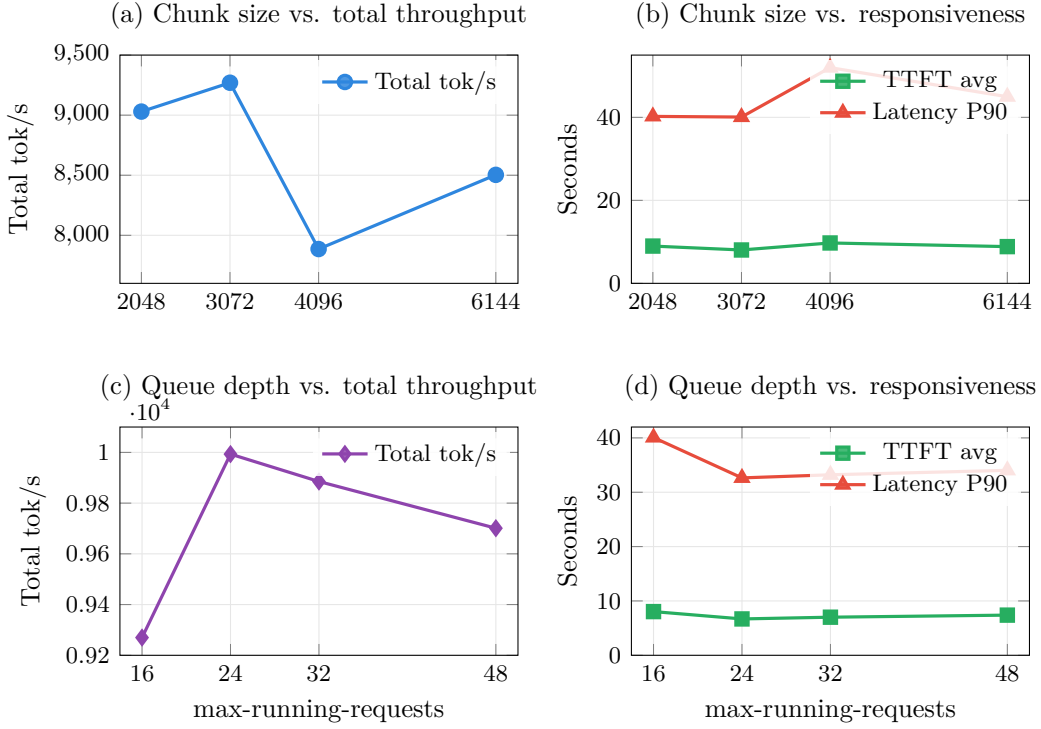
\begin{figure}[!htbp]
\centering
\begin{tikzpicture}
\begin{groupplot}[
  group style={group size=2 by 2, horizontal sep=1.5cm, vertical sep=1.9cm},
  width=0.43\textwidth,
  height=4.6cm,
  grid=both,
  major grid style={draw=gray!20},
  minor grid style={draw=gray!10},
  tick label style={font=\footnotesize},
  label style={font=\small},
  title style={font=\small},
  legend style={font=\footnotesize, draw=none, fill=white, fill opacity=0.85, text opacity=1},
]
\nextgroupplot[
  title={(a) Chunk size vs. total throughput},
  ylabel={Total tok/s},
  xmin=1800, xmax=6400,
  xtick={2048,3072,4096,6144},
  xticklabels={2048,3072,4096,6144},
  ymin=7600, ymax=9500,
]
\addplot[ccblue, line width=1.2pt, mark=*, mark size=2.5pt] coordinates {
  (2048,9029.64) (3072,9269.88) (4096,7885.54) (6144,8503.08)
};
\addlegendentry{Total tok/s}

\nextgroupplot[
  title={(b) Chunk size vs. responsiveness},
  ylabel={Seconds},
  xmin=1800, xmax=6400,
  xtick={2048,3072,4096,6144},
  xticklabels={2048,3072,4096,6144},
  ymin=0, ymax=55,
]
\addplot[ccgreen, line width=1.2pt, mark=square*, mark size=2.3pt] coordinates {
  (2048,8.98) (3072,8.04) (4096,9.73) (6144,8.85)
};
\addlegendentry{TTFT avg}
\addplot[ccred, line width=1.2pt, mark=triangle*, mark size=2.6pt] coordinates {
  (2048,40.23) (3072,40.07) (4096,51.98) (6144,44.97)
};
\addlegendentry{Latency P90}

\nextgroupplot[
  title={(c) Queue depth vs. total throughput},
  xlabel={max-running-requests},
  ylabel={Total tok/s},
  xmin=14, xmax=50,
  xtick={16,24,32,48},
  xticklabels={16,24,32,48},
  ymin=9200, ymax=10100,
]
\addplot[ccpurple, line width=1.2pt, mark=diamond*, mark size=2.5pt] coordinates {
  (16,9269.88) (24,9993.23) (32,9885.18) (48,9701.02)
};
\addlegendentry{Total tok/s}

\nextgroupplot[
  title={(d) Queue depth vs. responsiveness},
  xlabel={max-running-requests},
  ylabel={Seconds},
  xmin=14, xmax=50,
  xtick={16,24,32,48},
  xticklabels={16,24,32,48},
  ymin=0, ymax=42,
]
\addplot[ccgreen, line width=1.2pt, mark=square*, mark size=2.3pt] coordinates {
  (16,8.04) (24,6.69) (32,7.01) (48,7.39)
};
\addlegendentry{TTFT avg}
\addplot[ccred, line width=1.2pt, mark=triangle*, mark size=2.6pt] coordinates {
  (16,40.07) (24,32.64) (32,33.21) (48,34.03)
};
\addlegendentry{Latency P90}
\end{groupplot}
\end{tikzpicture}
\caption{Sensitivity of serving performance to chunked prefill size and queue depth under the OpenClaw long-context workload. Each plotted point corresponds to a measured configuration reported in Table~\ref{tab:full_sweep}. Among the tested chunk sizes, \texttt{3072} provides the best throughput--latency trade-off at \texttt{tp=4}, \texttt{pp=4}, and \texttt{max-run=16}. Among the tested queue-depth settings, \texttt{24} achieves the highest total token throughput while also yielding the lowest average TTFT and the best P90 latency.}
\label{fig:chunk_and_queue}
\end{figure}

\begin{figure}[!htbp]
\centering
\begin{tikzpicture}
\begin{groupplot}[
  group style={group size=2 by 1, horizontal sep=1.7cm},
  width=0.43\textwidth,
  height=5.2cm,
  ymajorgrids=true,
  major grid style={draw=gray!20},
  tick label style={font=\footnotesize},
  label style={font=\small},
  title style={font=\small},
  legend style={font=\footnotesize, draw=none, fill=white, fill opacity=0.85, text opacity=1},
]
\nextgroupplot[
  title={(a) TP/PP split at 3072 / max24},
  ybar,
  bar width=18pt,
  ymin=0.0, ymax=0.55,
  ylabel={Req/s},
  symbolic x coords={tp2-pp8,tp4-pp4,tp8-pp2},
  xtick=data,
  xticklabels={2/8,4/4,8/2},
  nodes near coords,
  every node near coord/.append style={font=\footnotesize, yshift=2pt},
]
\addplot[fill=ccblue!75, draw=ccblue!95!black] coordinates {
  (tp2-pp8,0.41) (tp4-pp4,0.48) (tp8-pp2,0.33)
};

\nextgroupplot[
  title={(b) Normalized serving cost under fixed hardware},
  ybar,
  bar width=12pt,
  ymin=0.75, ymax=1.20,
  ylabel={Normalized cost},
  symbolic x coords={baseline,legacy,recommended},
  xtick=data,
  xticklabels={2048/16,4096/16,3072/24},
]
\addplot[fill=ccorange!80, draw=ccorange!95!black] coordinates {
  (baseline,1.000) (legacy,1.132) (recommended,0.896)
};
\addlegendentry{Cost / request}
\addplot[fill=ccgreen!70, draw=ccgreen!70!black] coordinates {
  (baseline,1.000) (legacy,1.145) (recommended,0.904)
};
\addlegendentry{Cost / token}
\end{groupplot}
\end{tikzpicture}
\caption{Comparison of model-parallel partitioning efficiency and normalized serving cost. In the left panel, \texttt{tp=4}/\texttt{pp=4} delivers the highest request throughput among the tested TP/PP splits; bar labels report the measured req/s directly. In the right panel, normalized cost uses \texttt{2048/16} as the baseline (1.0) under the same hardware footprint, so lower is better. Relative to that baseline, the recommended \texttt{3072/24} profile reduces cost by about \CostReqVsBase{} per request and \CostTokVsBase{} per token; relative to the \texttt{4096/16} comparison profile, the reductions are about \CostReqVsLegacy{} and \CostTokVsLegacy{}, respectively.}
\label{fig:tppp_and_cost}
\end{figure}
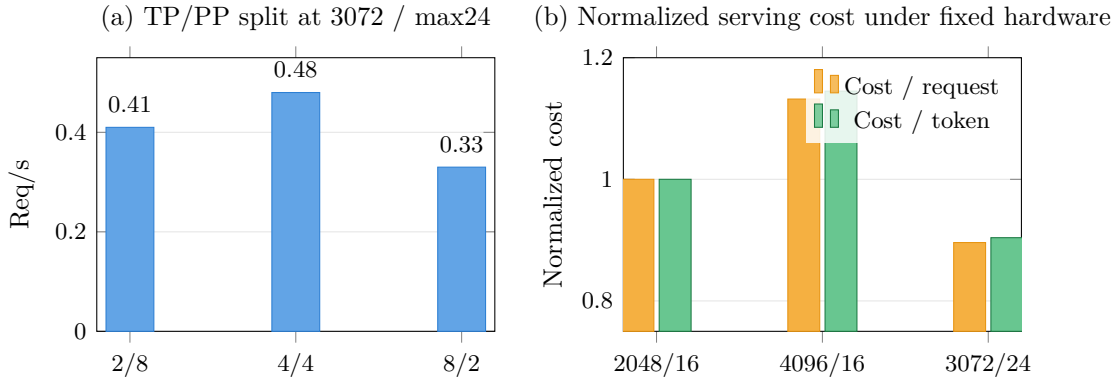

\textbf{Benchmark provenance and reproducibility scope.} The raw sweep values used in Table~\ref{tab:full_sweep}, Figure~\ref{fig:chunk_and_queue}, Figure~\ref{fig:tppp_and_cost}, and Table~\ref{tab:cost_summary} are taken from the internal GLM-5 serving stress-test log for the OpenClaw long-context workload~\citep{glm5_stress_test_internal}. Each row in the raw CSV corresponds to one aggregate \texttt{bench\_serving} measurement for a single server configuration. The available log records aggregate throughput and latency metrics but does not include repeated-run identifiers, a random seed, or a warm-up duration; consequently, this report does not claim confidence intervals or statistical significance. All recommendations are therefore phrased as \emph{best measured within the tested parameter space} under the fixed hardware and software environment, not as a universal optimum for GLM-5 serving.

\section{Experiments}
\label{sec:experiments}
\subsection{Experimental Setup}
This section describes the experimental platform, model artifact, and serving stack used for the parameter sweep.

\textbf{Testbed and hardware configuration.} All measurements were collected on the same distributed serving cluster so that parameter comparisons are made under a fixed hardware budget. The deployment platform consists of a \textbf{two-node, sixteen-GPU H100 cluster}, with \textbf{eight H100 GPUs per node}. Each server is equipped with \textbf{2$\times$ Intel Xeon Platinum 8468 CPUs} (48 cores per socket, for 96 physical cores and 192 hardware threads per node) and \textbf{2~TiB of system memory}. Local storage is organized as \textbf{2$\times$ 1.7~TB Samsung SATA SSDs in RAID1} for the system volume, together with \textbf{8$\times$ 7~TB Intel NVMe SSDs} for high-throughput data access and benchmark staging. The I/O and network inventory includes multiple \textbf{Mellanox ConnectX-7 InfiniBand adapters}, as well as \textbf{Intel E810-XXV} and \textbf{Mellanox ConnectX-6 Dx} Ethernet controllers. This hardware context is important for interpreting the parallelism study, because the tested \texttt{tp/pp} decompositions---\texttt{2/8}, \texttt{4/4}, and \texttt{8/2}---all consume the same sixteen accelerators but induce different collective-communication and pipeline-balance demands on the interconnect fabric \citep{vllm_parallelism_scaling}. By fixing CPU, memory, storage, and network configuration across the entire sweep, the study isolates the effect of serving parameters rather than mixing them with host-level resource variation.

\begin{table}[!htbp]
\centering
\scriptsize
\setlength{\tabcolsep}{4.0pt}
\caption{Measured parameter sweep for the OpenClaw long-context workload. Each raw CSV row is represented once; configuration is specified by the chunk/tp/pp/max-run columns, and the global-winner row also serves as the TP/PP reference point. Shaded rows mark the best tested point within a local sweep block or the global recommendation.}
\label{tab:full_sweep}
\resizebox{0.98\textwidth}{!}{%
\begin{tabular}{@{}lrrrrrrrrrrr@{}}
\toprule
Block / role & chunk & tp & pp & max-run & Out tok/s & Total tok/s & Req/s & TTFT avg & TTFT P90 & TPOT ms & Latency P90 \\
\midrule
Chunk & 2048 & 4 & 4 & 16 & 215.17 & 9029.64 & 0.43 & 8.98 & 14.18 & 54 & 40.23 \\
\rowcolor{blue!6}
Chunk best (max16) & 3072 & 4 & 4 & 16 & 220.89 & 9269.88 & 0.44 & 8.04 & 12.58 & 54 & 40.07 \\
Chunk & 4096 & 4 & 4 & 16 & 187.90 & 7885.54 & 0.38 & 9.73 & 17.39 & 63 & 51.98 \\
Chunk & 6144 & 4 & 4 & 16 & 202.62 & 8503.08 & 0.41 & 8.85 & 15.00 & 59 & 44.97 \\
\midrule
\rowcolor{green!10}
Global winner & 3072 & 4 & 4 & 24 & 238.13 & 9993.23 & 0.48 & 6.69 & 11.61 & 50 & 32.64 \\
Queue & 3072 & 4 & 4 & 32 & 235.55 & 9885.18 & 0.47 & 7.01 & 12.07 & 50 & 33.21 \\
Queue & 3072 & 4 & 4 & 48 & 231.16 & 9701.02 & 0.46 & 7.39 & 12.87 & 50 & 34.03 \\
4096/max32 check & 4096 & 4 & 4 & 32 & 225.47 & 9462.00 & 0.45 & 7.73 & 12.90 & 51 & 36.01 \\
\midrule
TP/PP & 3072 & 2 & 8 & 24 & 207.43 & 8705.06 & 0.41 & 7.80 & 12.96 & 57 & 37.41 \\
TP/PP & 3072 & 8 & 2 & 24 & 164.56 & 6905.99 & 0.33 & 9.48 & 16.50 & 73 & 47.55 \\
\bottomrule
\end{tabular}%
}
\end{table}

\textbf{Model artifact and precision scope.} The serving target is the \textbf{GLM-5 FP8 checkpoint}. Accordingly, all benchmark results in this report characterize \textbf{the FP8 deployment variant of GLM-5 on the current H100 cluster}. Quantization changes the model weight footprint and therefore the amount of GPU memory that remains available for long-context KV cache and for concurrent requests. It can also change runtime behavior indirectly through kernel selection, memory-fragmentation patterns, and communication pressure in distributed execution \citep{glm5_arxiv,glm5_vllm_ascend_doc}. For that reason, the recommended parameter setting derived in this report should not be over-generalized to BF16, INT4, W4A8, or other GLM-5 artifacts without separate measurement. The conclusions here are intentionally narrower: they identify the best \emph{tested} serving configuration for the FP8 OpenClaw deployment under the present hardware conditions.

\textbf{Serving stack and controlled variables.} The online service is implemented with \textbf{SGLang v0.5.9} in two-node distributed mode and uses the \texttt{flashinfer} attention backend. Across all experiments, the fixed server-side settings include \texttt{--nnodes 2}, \texttt{--served-model-name glm-5}, \texttt{--trust-remote-code}, \texttt{--tool-call-parser glm47}, \texttt{--reasoning-parser glm45}, \texttt{--enable-reasoning-tokens}, \texttt{--context-length 202752}, \texttt{--mem-fraction-static 0.85}, \texttt{--enable-metrics}, and \texttt{--enable-cache-report}. Only three serving knobs are intentionally varied in the sweep: \texttt{--chunked-prefill-size}, the tensor/pipeline parallelism split (\texttt{tp/pp}), and \texttt{--max-running-requests}. This experimental design provides a clean basis for comparison: the model artifact, cluster topology, and serving framework are held constant, so performance differences can be attributed directly to the scheduling and partitioning choices under study \citep{sglang_server_args}.

\section{Ablation Study}
\subsection{Chunked Prefill Size Ablation}
Long-prefix requests shift the compute balance toward the prefill stage: the server must process the entire prompt to build KV cache state before it can decode subsequent tokens. For tool-augmented agents, such long-prefill behavior is common because conversation history and tool outputs are injected back into context \citep{openclaw_context,glm5_arxiv}.

SGLang’s \texttt{--chunked-prefill-size} sets the maximum number of tokens per chunk during chunked prefill \citep{sglang_server_args}. In principle, a larger chunk can reduce scheduling overhead for very long prompts, but it can also make prefill bursts more monopolizing and worsen decode interference for other requests. A smaller chunk improves fairness, but may fragment long-prefix progress too aggressively.

The expanded sweep shows that the best measured point in this workload is \textbf{\texttt{chunked-prefill-size=3072}}, not 4096. Holding \texttt{tp=4}, \texttt{pp-size=4}, and \texttt{max-running-requests=16} fixed, moving from 2048 to 3072 raises total token throughput from \BaseTotalTokTput{} to 9269.88 tok/s and reduces TTFT average from \BaseTTFTAvg{} to 8.04 seconds, while keeping latency P90 nearly unchanged (\BaseLatPNineZero{} to 40.07 seconds). By contrast, 4096 and 6144 underperform in the max-16 chunk sweep, with 4096 being the weakest tested chunk size on both throughput and latency. A high-chunk cross-check at \texttt{4096/4/4/32} improves over \texttt{4096/4/4/16}, but it still remains below the recommended \texttt{3072/4/4/24} point on request throughput, total-token throughput, TTFT average, and latency P90.

\textbf{Interpretation.} For the current OpenClaw traffic pattern, 2048 appears slightly too conservative for long-prefix efficiency, while 4096 and above are already large enough to introduce chunk-level burstiness without compensating gains. The sweet spot therefore sits in the middle: 3072 advances long prefills more efficiently than 2048 while avoiding the latency penalty observed at 4096 and 6144 \citep{sglang_server_args}.

\begin{table}[!htbp]
\centering
\footnotesize
\setlength{\tabcolsep}{3pt}
\caption{Cost-oriented deployment summary under a fixed hardware footprint. Normalized cost is defined as inverse throughput relative to the 2048/16 baseline.}
\label{tab:cost_summary}
\begin{tabular}{@{}lrrrrrrrrl@{}}
\toprule
Profile & chunk & tp & pp & max-run & Req/s & Total tok/s & Cost / req & Cost / tok & Decision note \\
\midrule
Baseline 2048/16 & \BaseChunk & \BaseTP & \BasePP & \BaseMaxRun & \BaseReqTput & \BaseTotalTokTput & \NormCostReqBase & \NormCostTokBase & Rollback point \\
Comparison 4096/16 & \LegacyChunk & \LegacyTP & \LegacyPP & \LegacyMaxRun & \LegacyReqTput & \LegacyTotalTokTput & \NormCostReqLegacy & \NormCostTokLegacy & Not recommended \\
\rowcolor{green!10}
Recommended 3072/24 & \RecChunk & \RecTP & \RecPP & \RecMaxRun & \RecReqTput & \RecTotalTokTput & \NormCostReqRec & \NormCostTokRec & Default profile \\
\bottomrule
\end{tabular}
\end{table}

\subsection{TP/PP Parallelism Ablation}
SGLang supports both tensor parallelism (\texttt{--tp-size}) and pipeline parallelism (\texttt{--pp-size}) \citep{sglang_server_args}. Tensor parallelism splits intra-layer computation and introduces collective communication; pipeline parallelism splits layers across stages and can suffer stage imbalance and bubble overhead \citep{megatron_lm,gpipe}. vLLM documentation similarly notes that distributed serving is highly sensitive to the TP/PP split and interconnect properties \citep{vllm_parallelism_scaling}.

The new data preserve the earlier direction on this axis: \textbf{\texttt{tp=4} and \texttt{pp-size=4} is still the best tested split}. Under the same \texttt{chunked-prefill-size=3072} and \texttt{max-running-requests=24}, the 4/4 split reaches \RecReqTput{} req/s and \RecTotalTokTput{} tok/s, compared with 0.41 req/s and 8705.06 tok/s for 2/8, and 0.33 req/s and 6905.99 tok/s for 8/2. TTFT average is also best at 6.69 seconds, versus 7.80 seconds for 2/8 and 9.48 seconds for 8/2.

\textbf{Interpretation.} In this cluster, \texttt{tp=4} / \texttt{pp=4} appears to strike the right balance among three competing effects: per-device memory headroom for long-context KV cache, communication cost for TP collectives, and pipeline balance for PP stages. The 2/8 split likely pays more pipeline bubble cost, while the 8/2 split likely pays more communication and synchronization overhead. As before, this conclusion should be stated conservatively as \emph{best among the tested splits on the current hardware topology} \citep{vllm_parallelism_scaling}.

\subsection{Max-Running-Requests Ablation}
SGLang’s \texttt{--max-running-requests} caps the number of simultaneously running requests \citep{sglang_server_args}. In many serving setups, raising this cap increases effective batch size and throughput but eventually worsens queueing delay and tail latency \citep{sglang_hparam_tuning}.

The best measured point in this sweep is \textbf{\texttt{max-running-requests=24}}, not 16 or 32. Holding \texttt{chunked-prefill-size=3072}, \texttt{tp=4}, and \texttt{pp-size=4} fixed, moving from 16 to 24 increases request throughput from 0.44 to \RecReqTput{} req/s and total token throughput from 9269.88 to \RecTotalTokTput{} tok/s. At the same time, TTFT average drops from 8.04 to \RecTTFTAvg{} seconds and latency P90 drops from 40.07 to \RecLatPNineZero{} seconds. Thus, 24 is not merely a higher-throughput mode; it is a strict improvement over 16 on both efficiency and responsiveness in this workload.

Beyond 24, the gains flatten and then reverse in the 3072 queue sweep. At 32 and 48, throughput remains below the 24-run peak, while TTFT and latency P90 rebound slightly. The separate \texttt{4096/4/4/32} cross-check confirms that simply pairing a larger chunk with deeper queueing does not surpass the 3072/max24 profile. This suggests that the concurrency knee of the current OpenClaw workload sits around 24 running requests. Past that point, queueing and scheduling overhead begin to dominate the marginal batching benefit.

For this deployment, \texttt{max-running-requests=24} is the best tested queue-depth setting. It combines the highest measured throughput with the best TTFT and latency P90 among the evaluated alternatives in the current environment \citep{sglang_hparam_tuning,vllm_anatomy_metrics}.

\paragraph{Cost reduction under a fixed hardware footprint.}
Because all configurations in the current sweep use the same model and the same serving footprint, normalized serving cost can be estimated as the inverse of achieved throughput. On a request basis,
\[
\mathrm{NormCost}_{\mathrm{req}}(i)=\frac{\mathrm{RPS}_{\mathrm{baseline}}}{\mathrm{RPS}_{i}},
\]
and on a token basis,
\[
\mathrm{NormCost}_{\mathrm{tok}}(i)=\frac{\mathrm{TotalTok/s}_{\mathrm{baseline}}}{\mathrm{TotalTok/s}_{i}}.
\]
Applying these definitions to the measured winner \texttt{3072/4/4/24} yields a normalized cost of \NormCostReqRec{} per request and \NormCostTokRec{} per token, which means a reduction of \CostReqVsBase{} and \CostTokVsBase{} relative to the \texttt{2048/4/4/16} baseline. Against the \texttt{4096/4/4/16} comparison profile, the savings are even larger: \CostReqVsLegacy{} per request and \CostTokVsLegacy{} per token.

\FloatBarrier
\section{Conclusions and Recommendations}

\textbf{Primary conclusion.} Within the measured parameter space, the best tested configuration for OpenClaw-serving GLM-5 is:

\begin{center}
\small
\begin{tabular}{@{}rl@{}}
\texttt{chunked-prefill-size} & \texttt{3072} \\
\texttt{tp / pp-size} & \texttt{4 / 4} \\
\texttt{max-running-requests} & \texttt{24} \\
\end{tabular}
\end{center}

This profile is the measured winner on both throughput and latency among the tested candidates, rather than a compromise that sacrifices responsiveness for throughput.

\textbf{Supporting evidence.} The measured results indicate: (i) 3072 is a better chunk size than both 2048 and 4096 for this long-context workload, (ii) the queue-depth optimum sits at 24 rather than 16 or 32, and (iii) the 4/4 TP/PP split remains the best tested partitioning choice.

\textbf{Quantified business impact.} Relative to the conservative \texttt{2048/4/4/16} baseline, the recommended profile improves request throughput by \GainReqVsBase{} and total token throughput by \GainTotVsBase{}, while reducing TTFT average by \DropTTFTVsBase{} and latency P90 by \DropLatVsBase{}. Under constant hardware, this translates into an estimated serving-cost reduction of \CostReqVsBase{} per request and \CostTokVsBase{} per token. Relative to the \texttt{4096/4/4/16} comparison profile, the recommended setting reduces normalized serving cost by roughly \CostReqVsLegacy{} per request and \CostTokVsLegacy{} per token.

\textbf{Operational recommendation.} Use \texttt{3072 / tp4 / pp4 / max24} as the default OpenClaw profile. Keep \texttt{2048 / 4 / 4 / 16} as a conservative rollback point during production rollout if needed. Treat \texttt{4096 / 4 / 4 / 16} and TP/PP splits such as \texttt{2/8} or \texttt{8/2} as not recommended for the current workload, because they are strictly worse on the measured axes.

\textbf{Further work.} A subsequent round should add resource telemetry (GPU utilization, memory headroom, KV-cache occupancy), archive exact benchmark command lines, and repeat the same sweep under several traffic mixes such as pure chat, retrieval-heavy, and tool-heavy sessions. For the present internal deployment decision, the existing measurements are sufficient to nominate the best tested parameter combination and quantify the associated serving-cost reduction; stronger external performance claims should be backed by repeated runs and confidence intervals.

\FloatBarrier
{\small
\begingroup
\sloppy
\bibliographystyle{ieee}
\bibliography{egbib}
\endgroup
}

\end{document}